\documentclass[twocolumn,aps]{revtex4}
\usepackage{graphicx}  
\usepackage{dcolumn}   
\usepackage{bm}        
\usepackage{amssymb}   
\begin{document}

\title{Free fall onto evaporating black holes at the quantum limit}
\author{Maurice H.P.M. van Putten}
\affiliation{Department of Astronomy and Space Science, Sejong University, Gwangjin-gu, Seoul 143-747, Korea}

\def\runningauthor{DRAFT V1.0}
\def\runningtitle{holography, computation}

\begin{abstract}
Black hole space times evaporate in discrete steps due to remarkably slow Hawking radiation. We here
identify evaporation with essentially extremal states at the limit of quantum computation,
performing $2.7\times 10^{79}$ bit calculations per photon emission in a one solar mass black hole. During evaporation, particles in free fall co-evolve satisfying $EM=$constant, where $E$ and $M$ denote the 
total mass energy-at-infinity of the particle and, respectively, black hole. Particles are hereby increasingly 
entangled with the black hole space-time over the course of its evaporation.
\end{abstract}

\pacs{black holes: information, computation}

\maketitle

\section{Introduction\label{sec:intro}}

The evaporation of black holes is a singular perturbation away from thermal equilibrium described by power laws in luminosity $L$ and temperature $T$ as a function of total mass energy-at-infinity $Mc^2$, where $c$ denotes the velocity of light, i.e., 
\begin{eqnarray}
L \propto M^{-p},~~T\propto M^{-q}
\label{EQN_LT}
\end{eqnarray}
with $p,q>0$ that is characteristic for open self-gravitating systems with negative specific heat. Evaporation by Hawking radiation \citep{haw77} shows $p=2$ and $q=1$. Classical systems such as globular clusters 
are qualitatively similar with $p\simeq 0.9$ and $q\simeq 6$ \citep{van12}. For
Hawking radiation \citep{haw77}, emission of individual photons is once
every few thousand light crossing time scales \cite[][and below]{bek95,hod15}. A
Schwarzschild given by the line-element 
\begin{eqnarray}
ds^2=-\alpha^2 dt^2 + \frac{dr^2}{\alpha^2} + r^2\left( d\theta^2 + \sin^2\theta d\varphi^2\right),
\label{EQN_Schw}
\end{eqnarray}
hereby changes in essentially discrete steps by individual photon emissions one-by-one
\begin{eqnarray}
\dot{M}(t) =- \sum_{i=1}^N \epsilon_i \delta(t-t_i)
\label{EQN_Md}
\end{eqnarray}
at instances $t_i$, where $\epsilon_i$ denotes the energy of the photon emitted measured at infinity, 
$\delta(t)$ is the Kronecker delta function, $\alpha = \sqrt{1 - 2R_g/r}$ is the redshift factor,
$R_g=2GM/rc^2$ denotes the gravitational radius, $G$ is Newton's constant and $c$ is the velocity of light. 

As solutions to classical general relativity, black holes hide 
astronomical amounts of information given by the Bekenstein-Hawking entropy 
\citep{bek73,haw77,hod98} 
\begin{eqnarray}
k_B^{-1} S=\frac{1}{4}A_Hl_p^{-2}, ~~A_H=4 \ln3 \,n\, l_p^2
\label{EQN_0}
\end{eqnarray}
in $n$ Planck sized surface elements $l_p^2$ for a black hole surface area $A_H=16\pi R_g^2$, where $l_p=\sqrt{G\hbar/c^3}$ denotes the Planck length in terms of Planck's constant $\hbar$ and the Boltzmann constant $k_B$. 

In the $S$-matrix formalism (e.g. \citep{ste94}), the complete evaporation of the black hole represents a unitary
evolution. It implies that all of $S$ in (\ref{EQN_0}) is ultimately recovered in information projected onto the celestial sphere by Hawking radiation, at times following the evaporative lifetime $t_{ev}$ of the black hole. In the
continuum approximation, the black hole evolution satisfies the finite-time singularity  
\begin{eqnarray}
M(t) = M_0 \left( 1- \frac{t}{t_{ev}}\right)^{\frac{1}{3}}
\label{EQN_Mt}
\end{eqnarray}
for an initial mass $M_0$ at the Hawking evaporation time 
\begin{eqnarray}
t_{ev} = 1280\pi\, t_c\, k_B^{-1} \,S,
\label{EQN_tev}
\end{eqnarray}
where $t_c={R_g}/{c}$ denotes the light crossing time scale of the black hole.
In (\ref{EQN_tev}), $t_{ev}$ scales with $k_B^{-1}S$, giving the well-known astronomically
large time scales for macroscopic black holes. (For globular clusters, $t_{ev}$ scales with
the relaxation time, giving typical scales on the scale of the Hubble time.)

\begin{figure}[h]
\centerline{\includegraphics[width=80mm,height=60mm]{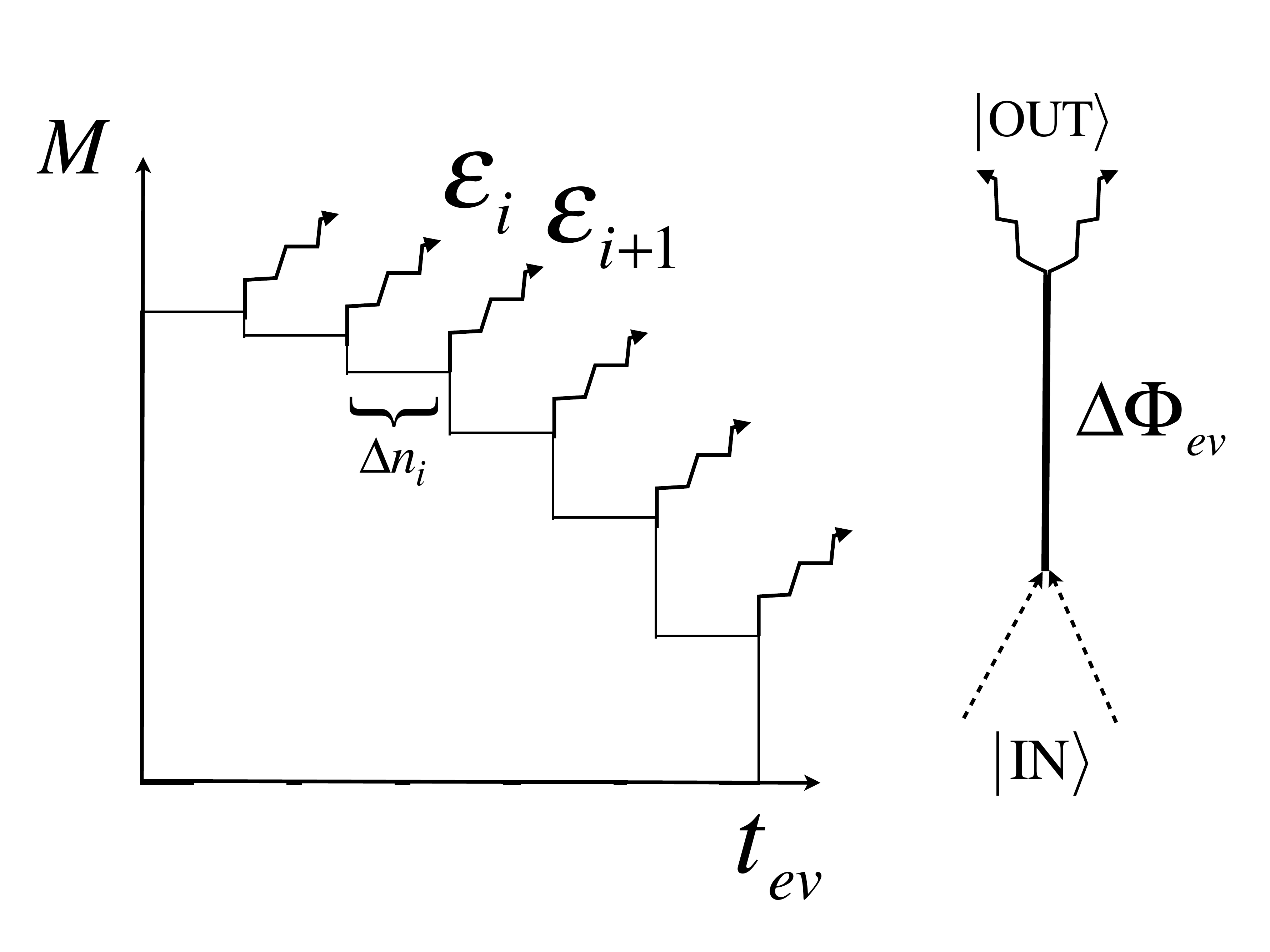}}
\caption{Schematic overview of the discrete evolution of a Schwarzschild  
black hole space-time of mass $M$ from sequences of computations $\Delta n_i$
each leading to a photon emission. The cumulative result is a finite 
phase change $\Delta \Phi_{ev}\simeq Mc^2 t_{ev}/\hbar$ over its evaporative lifetime $t_{ev}$.}
\label{fig1}
\end{figure}

Here, we shall identify the exceedingly slow and discrete evaporation process
with a strict limit of quantum computation in unitary evaporation in the $S$-matrix
formalism $\left| \mbox{OUT}  \right. > = \hat{S}\left|\mbox{ IN} \right.>$  (Fig. \ref{fig1}),
wherein 
\begin{eqnarray}
\dim \hat{S}  = \left( k_B^{-1} S \right) \times \left( k_B^{-1} S\right)
\label{EQN_dimS}
\end{eqnarray}
represents the total number of calculations over the course of complete evaporation,
producing state in remnant Hawking radiation recovering the information (\ref{EQN_0}).
In this process, the rate of computations is bounded by 4 times the angular velocity 
of the wave function of the black hole space time, defined by its mass $M(t)$ \citep{mar98}
\begin{eqnarray}
\dot{n}=\frac{2Mc^2}{\pi\hbar} = 1.09\times 10^{81}\left(\frac{M}{M_\odot}\right) \,\mbox{bit~s}^{-1}
\label{EQN_dn}
\end{eqnarray}
with an associated total computation of 
\begin{eqnarray}
\frac{2}{\pi} \Delta \Phi_{ev}\simeq  t_{ev} \dot{n}\,\mbox{ bits}.
\label{EQN_Phi}
\end{eqnarray}
Thus, {\em unitarity prohibits arbitrarily fast evaporation}.

To probe of the evolution of the black hole during evaporation, we consider a perturbation by dropping
in a particle of mass $E$ (posterior to the black hole formation). While an observer in free
fall onto a black hole is not aware of an event horizon nor, probably, any associated Hawking radiation,
it is aware of a time-dependent tidal field. Defined by the square root of the Kretschmann invariant \citep{hen00} 
\begin{eqnarray}
Q = 4\sqrt{3}\frac{R_g}{r^3},
\label{EQN_Q}
\end{eqnarray}
$Q$ is a scalar field that may invariably be parameterized by the observers' eigentime $\tau$ or time-at-infinity $t$. 
An observer in free fall is hereby aware of $M$ upon measuring $Q$. Observers ${\cal O}^\prime$ 
at fixed Schwarzschild radius with zero angular momentum (ZAMO's \citep{tho86}) 
notice $M$ evolving with respect to their eigentime $\tau$ according to
$\left( {dM}/{d\tau}\right)_{ZAMO} = - {L}/{\alpha}.$
Observers ${\cal O}$ in free fall detect $dM/d\tau$ be greater than the ZAMO rate by an additional Lorentz factor. To be sure, ${\cal O}$ hereby concludes that $M$ is rapidly evolving, not constant as is customary
assumed (see also \citep{vac07} for a discussion).
To evaluate the evolution of the Schwarzschild space-time observed in free fall,
we set out to solve for the geodesic trajectories, taking into account jump conditions associated with the discrete evolution sketched in Fig. \ref{fig1}. We hereby shall identify $M(\tau)$ and a correlation with ${\cal O}$'s  total energy at infinity $E(\tau)$.

n \S2, we identify $t_{ev}$ with extremal quantum computation in unitary evaporation. During
evaporation, we solve for jump conditions in geodesic motion of massive observers subject to (\ref{EQN_Md})
in \S3, that are summarized by a new conserved quantity that expresses a co-evolution of $E$ and $M$
(\S4). In \S5, we give some concluding remarks on entanglement of ${\cal O}$ with the black hole
space time.

\section{Extremal quantum computation}

The energy spectrum of Hawking radiation photons is effectively that of a black body at the finite horizon temperature
\begin{eqnarray}
k_B T_H = \frac{\hbar c}{8\pi R_g},
\label{EQN_1a}
\end{eqnarray}
giving an average photon energy $\epsilon_\gamma=Ck_BT_H$. For
black body radiation, Planck's law $u=(2hc^{-2}\nu^3/(e^x-1)$, $x=h\nu/k_BT$, for the energy
density in radiation per unit solid angle and frequency $\nu$ implies
\begin{eqnarray}
C=\frac{f(3)}{f(2)},~~f(p)=\sum_{0}^\infty \frac{x_n^p}{e^{x_n}-1}\Delta x,
\label{EQN_C}
\end{eqnarray}
where $x_n=n\Delta x$ and $\Delta x=h\Delta \nu/k_BT$ denotes a spacing in dimensionless frequency.
In the continuum limit obtain $C=2.7012$. However, the temperature (\ref{EQN_1a}) represents an 
energy level spacing defined by the horizon quantization (\ref{EQN_0}). 
With $h\Delta \nu = k_BT\log3$, $C=2.7955$. The photon emission rate $\dot{N}=L/\epsilon_\gamma$ 
associated with the luminosity $L=\hbar c^2/15360\pi R_g^2$ of black body radiation at (\ref{EQN_1a}), satisfies
\begin{eqnarray}
\dot{N}_\gamma = \frac{1}{1920C} t_c^{-1} \simeq \frac{1}{5300} t_c^{-1},
\label{EQN_1c}
\end{eqnarray}
where $t_c=R_g/c$ denotes the light crossing time. This emission rate of one photon per few thousand
light crossing time scales is a key property of Hawking radiation, previously identified in \citep{bek95}. 
This emission frequency satisfies
\begin{eqnarray}
\nu \simeq 40\left(\frac{M}{M_\odot}\right)^{-1}\mbox{Hz}.
\label{EQN_nu}
\end{eqnarray}
The estimates (\ref{EQN_1c}-\ref{EQN_nu}) show a discrete evolution (\ref{EQN_Md}) 
in steps of size $\epsilon=Ck_BT$.

If each photon carries of $k$ bits to infinity, the luminisotiy $L=-\dot{M}c^2$, the total amount of information received is
\begin{eqnarray}
I = k\log2\int_0^{t_{ev}} \dot{N}dt = \frac{k\log 2}{C} \int_0^M \frac{c^2dM}{\epsilon_\gamma} \simeq \frac{k}{4} k_B^{-1}S
\label{EQN_2}
\end{eqnarray} 
within 1\% in the above mentioned approximation $C = 2.7955$.
Our information extraction scheme hereby follows the power law
\begin{eqnarray}
I(t)\simeq \frac{k}{4}k_B^{-1} S \left(\frac{t}{t_{ev}}\right)^\frac{2}{3}~~(0\le t \le t_{ev})
\label{EQN_It}
\end{eqnarray}
to within the same precision.

During evaporation, the black hole
is susceptible to small kicks received in the process of emitting each photon. As a result,
the net momentum the black hole obtains from
a random walk in momentum space. Let $\delta \alpha= \epsilon_\gamma/Mc^2$ denote the angular perturbation
introduced by emitting one photon. Over the course of complete evaporation, it grows by a random walk to 
about $\Delta \alpha = \sqrt{N}\, \delta\alpha$, i.e., 
\begin{eqnarray}
\Delta \alpha \simeq \frac{A_n}{l_p^2} \times \frac{\hbar c}{8\pi R_g Mc^2} \simeq 2. 
\label{EQN_5}
\end{eqnarray}
A $k=4$ encoding \citep[cf.][]{van15b}.  comprising polarization and direction (in photon momenta) would hereby be
effective over the lifetime of the black hole, thus allowing in principle all information
in (\ref{EQN_0}) to be projected onto the celestial sphere. 

In unitary evaporation with information recovery according to (\ref{EQN_2}), $t_{ev}$ is bounded below by (\ref{EQN_dimS}-\ref{EQN_dn}),
\begin{eqnarray}
t_{ev} \simeq \frac{\dim \hat{S}}{\dot{n}} = 2\pi t_c k_B^{-1} S.
\label{EQN_4}
\end{eqnarray}
This bound explains the long duration (\ref{EQN_tev}). Specifically, the number of calculations
involved in preparing for a photon emission satisfies
\begin{eqnarray}
\Delta n = \frac{ \dot{n}}{\dot{N}_\gamma} =  \frac{960C}{\pi^2} k_B^{-1}S =2.7 \times 10^{79} \left(\frac{M}{M_\odot}\right)\,\mbox{bits}.
\label{EQN_Dn}
\end{eqnarray}

In the above, we have neglected information in any accompanying Hawking radiation in gravitational waves,
as its luminosity is mere 1\% of that in electromagnetic radiation \citep{pag13}. (Also, gravitational waves are
notoriously difficult to detect.) Conceivably, small perturbations
such as by these to the evolution of the black hole may account for our estimate $\alpha$ in (\ref{EQN_2}) being slightly
less than one by a similar amount. Furthermore, as pointed out in \cite{hod98}, surface discretization of the horizon area points to Shannon entropy in units of $\log 3\,l_p^2$, where $\log 3$ refers to three states in each Planck sized unit. The relation of the third mode, beyond the two in the present 4 bit-encoding, remains elusive, perhaps highlighting
our lack of understanding of the full quantum mechanical Hamiltonian describing the evolution of the black hole
space-time during its evaporation. Nevertheless, the overall agreement within about one percent between 
information extraction by Hawking radiation and entropy in the event horizon suggests that electromagnetic
radiation alone essentially accounts for unitary evolution. 

\section{Jump conditions in free fall}

In what follows, we use geometrical units $(G=c=1$), whereby $R_g=M$.
We assume that the observer ${\cal O}$ has a total energy satisfying the {accretion condition} 
$k_BT < E << M.$  
In the Schwarzschild line-element, ${\cal O}$'s velocity along a radial trajectory satisfies
\begin{eqnarray}
u^b=(\dot{t},\dot{r},0,0)= \left( \alpha^{-1} \cosh\mu,-\alpha \sinh\mu,0,0\right),
\label{EQN_F2}
\end{eqnarray} 
parametrized by a rapidity $\mu$ satisfying the normalization $u^cu_c=-1$. Here,
$u^r<0$ $(\mu>0)$ refers to radial infall. Here, $u^t=dt/d\tau$ combines the
combined result of gravitational redshift and Lorentz time-dilation, associated 
with the Lorentz factor along the trajectory.

In between successive photon emission events, $M$ is constant. Over the course of 
a time interval $N_\gamma^{-1}$, the redshift factor reduces according to
\begin{eqnarray}
\alpha^2=e^{-\lambda} \sim e^{-\frac{\Delta t}{2(M+E)}}\sim e^{-2650},
\label{EQN_exp}
\end{eqnarray}
i.e., zero in any quantization scheme of particle trajectories. Between any
two photon emissions, therefore, ${\cal O}$ has completely settled down to $r\simeq 2(M+E)$ as viewed from infinity. 
This asymptotic behavior hereby represents a singular limit for $u^t$.
Absent photon emissions, $(\partial_t)^b=(1,0,0,0)$ is a Killing vector, and the four-momentum $p^b=mu^b$ satisfies conservation of total energy-at-infinity 
\begin{eqnarray}
E=m\eta = m\alpha^2 u^t  = m\alpha\cosh\mu.
\label{EQN_e1}
\end{eqnarray}

The normalization condition $u^cu_c=-1$ gives $\dot{r}^2 = \eta^2-\alpha^2 \simeq \eta^2$
in the saturation limit when $\alpha^2$ is close to zero, whereby 
\begin{eqnarray}
2\dot{M}\simeq\dot{r}\simeq -\eta.
\label{EQN_rdot}
\end{eqnarray}
With $dM/d\tau=u^t dM/dt=-Lu^t$, it implies a finite $u^t=dt/d\tau$, i.e.,
\begin{eqnarray}
\frac{d\tau}{dt}\simeq \frac{2L}{\eta}.
\label{EQN_alpha2}
\end{eqnarray}
The limit $L=0$ recovers the conventional limit $d\tau/dt=0$ for a test particle falling onto
a classical Schwarzschild black hole. 

According to the geodesic equations of motion $u^c\nabla_c u^b=0$, 
$d u^b/d\tau + \Gamma^b_{ac}u^au^c=0.$ Between photon emissions $t_i < t < t_{i+1}$, 
we retain Christoffel symbols containing the impulsive changes $\delta M$, given by
\begin{eqnarray}
\Gamma^{t}_{tt} = \frac{\dot{M}}{r-2M}, \Gamma^{t}_{rr}=\frac{\dot{M}r^2}{(r-2M)^3}, \Gamma^{r}_{rt}=\frac{\dot{M}}{r-2M}.
\end{eqnarray} 
With $du^b/d\tau=u^t(du^b/dt)$, we obtain 
\begin{eqnarray}
\left(u^t\right)^{-1} \frac{du^t}{dt} + \Gamma^t_{tt} + \Gamma^t_{rr} \left(\frac{u^r}{u^t}\right)^2  \simeq 0,\\
\left(u^r\right)^{-1} \frac{d u^r}{dt} + 2 \Gamma^r_{rt}   \simeq 0.
\end{eqnarray} 
With $\dot{M}\delta t=-\epsilon$ and
\begin{eqnarray}
 \frac{\delta u^t}{u^t}  \simeq \frac{\delta\eta}{\eta} + e^{-\lambda}\delta e^\lambda,~~
 \frac{\delta u^r}{u^r} \simeq \frac{  \delta \eta}{\eta} + \frac{2}{\sinh(2\mu)}\delta \mu,
 \label{EQN_E2}
 \end{eqnarray}  
 we arrive at ${\delta\eta}/{\eta} + e^{-\lambda}\delta e^\lambda \simeq ({\epsilon}/{r \alpha^2})\left( 1+\tanh^2\mu\right),$
i.e.,
\begin{eqnarray}
\frac{\delta\eta}{\eta} + e^{-\lambda} \delta e^\lambda \simeq \frac{\epsilon}{M}\left(\frac{1+\tanh^2\mu}{2}\right)e^{\lambda}.
\label{EQN_E2a}
\end{eqnarray}
Similarly, 
${ \delta \eta}/{\eta}+ {2\delta\mu}/{\sinh(2\mu)} \simeq  ({2\epsilon}/{r})$ gives
\begin{eqnarray}
 \frac{  \delta \eta}{\eta}+  \frac{2\delta \mu}{\sinh(2\mu)} \simeq  \frac{\epsilon}{M}.
 \label{EQN_E2b}
 \end{eqnarray}  
 In the ultra-relativistic limit (large $\mu$), (\ref{EQN_E2a}) and (\ref{EQN_E2b}) reduce to
 \begin{eqnarray}
 \frac{\delta\eta}{\eta} + e^{-\lambda}\delta e^\lambda \simeq \frac{\epsilon}{M}e^{\lambda},~~
 \frac{  \delta \eta}{\eta} \simeq  \frac{\epsilon}{M}.
 \label{EQN_E2d}
 \label{EQN_E2c}
 \end{eqnarray}
  
\section{$EM$ is constant}

The ultra-relativistic limit (\ref{EQN_E2d}) relates $(\delta\eta,\delta M)$ across instances
of photon emission. With $\epsilon = - \delta M$ and constant $m$, (\ref{EQN_E2d}) gives 
the integral of motion
\begin{eqnarray}
E M \simeq m j
\label{EQN_j}
\end{eqnarray}
for some constant $j$. By (\ref{EQN_rdot}-\ref{EQN_alpha2}), we have
\begin{eqnarray}
3kt_{ev} = M_0^3,
\label{EQN_k}
\end{eqnarray}
where $k$ is short-hand in the black hole luminosity $dM/dt = -k M^{-2}$. 
Hence, (\ref{EQN_j}) gives $j = ({2k}/{M}) u^t.$ By (\ref{EQN_rdot}) and (\ref{EQN_j}), furthermore,
\begin{eqnarray}
M(\tau) \simeq M_0 \left( 1 - \frac{j\tau }{M_0^2}\right)^\frac{1}{2},
\label{EQN_E2e}
\end{eqnarray}
establishing the evolution of $M$ observed by ${\cal O}$. 

The finite value (\ref{EQN_alpha2}) of the correlation $u^t$ between $t$ and ${\cal O}$'s eigentime
can be further expressed in terms of a finite distance to the event horizon, as seen in the
Schwarzschild line-element of a ZAMO. By differentiation of (\ref{EQN_E2e}),
\begin{eqnarray}
\frac{t}{t_{ev}} \simeq 1 - \left( 1- \frac{j\tau}{M_0^2}\right)^{\frac{3}{2}}.
\label{EQN_tt}
\end{eqnarray}
we have
\begin{eqnarray}
u^t = \frac{3}{2}\left(\frac{t_{ev}}{M_0}\right)\left(\frac{M}{M_0}\right) \frac{j}{M_0}.
\end{eqnarray}
According to (\ref{EQN_e1}), we have $u^t\simeq (E/m)e^\lambda$, so that
\begin{eqnarray}
e^\lambda\simeq \left(\frac{m}{E} \right) u^t = \frac{3}{2} \left(\frac{t_{ev}}{M_0}\right)
\left(\frac{M}{M_0}\right)^2.
\end{eqnarray}
That is 
\begin{eqnarray}
e^\lambda\simeq \frac{M^2}{2k} = -\frac{1}{2}\left(\frac{dM}{dt}\right)^{-1}
= 960C\left( \frac{M}{\epsilon}\right),
\label{EQN_960}
\end{eqnarray}
where we used (\ref{EQN_1c}) and (\ref{EQN_k}). It follows that (\ref{EQN_960}) implies
\begin{eqnarray}
\delta \lambda = \ln\left(960 C+1\right)\simeq 8.58
\label{EQN_dla}
\end{eqnarray}
at each photon emission. During the (on average) time interval $\Delta t = 1920CM$ between
photon emissions, this shift (\ref{EQN_dla}) relaxes again to (\ref{EQN_960}). This
result is insensitive to $\delta U$. 

\section{Conclusions}

Hawking radiation is identified with discrete evolution of a Schwarzschild space time, operating
essentially at the quantum computation limit allowed by its total mass-energy at infinity. 
In particular, each photon emission results from an astronomically large calculation that,
accumulated over all photon emissions, results in a total calculation quantitatively in 
agreement with a unitary $S$-matrix formulation of complete black hole evaporation.  

The { discrete} evolution (\ref{EQN_Md}) of space-time with long time intervals $t_i < t < t_{i+1}$
of static space-time extending over a few thousand light crossing time scales enables us to 
calculate in detail the trajectory of an observer in free fall by integration of the geodesic equations of 
motion. The result shows a correlation (\ref{EQN_j}) between ${\cal O}$'s total mass energy $E$ and $M$,
which reduces to (\ref{EQN_j}). 

In a holographic interpretation \citep{van15b}, (\ref{EQN_0}) is the information on screens enveloping
the event horizon, encoding the position of a central singularity of mass $M$ at a radial distance 
$R_S=2R_g$. (The appearance of the event horizon is finite temperature effect by acceleration of the
screen, positioned at the constant Schwarzschild radius $R_S$. At the zero temperature of a
screen in free fall, such horizon is absent.) A mass-energy $E<<M$ falling in adds to $S$ an
additional information $8\pi EM$. By (\ref{EQN_j}), it is conserved during black hole evaporation, as long as the accretion condition $E>k_B T$ is satisfied.

{\bf Acknowledgments.} This work was partially supported by a Sejong University Faculty Research fund. The
author gratefully acknowledges stimulating discussions with S. Kawai, J. Polchinski, and 
G. 't Hooft on an earlier version of this manuscript.

\end{document}